\documentclass[hidelinks,journal,onecolumn]{IEEEtran}
\usepackage{amsmath,amsfonts,amssymb}
\usepackage{algorithmic}
\usepackage{algorithm}
\usepackage{array}
\usepackage[caption=false,font=footnotesize]{subfig}
\usepackage{textcomp}
\usepackage{stfloats}
\usepackage{url}
\usepackage{verbatim}
\usepackage{graphicx}
\usepackage{cite}
\usepackage{orcidlink}
\usepackage{lipsum}
\usepackage{outlines}
\usepackage{soul}
\usepackage[acronym,shortcuts]{glossaries}

\newacronym{AoD}{AoD}{angle of departure}
\newacronym{AoA}{AoA}{angle of arrival}
\newacronym{OS-QSM}{OS-QSM}{optimised scalable QSM}
\newacronym{SCCR}{SCCR}{sparse codesigned communication and radar}
\newacronym{GSM}{GSM}{generalised spatial modulation}
\newacronym{FDFR}{FDFR}{full-diversity full-rate}
\newacronym{mMIMO}{mMIMO}{massive multiple-input multiple-output}
\newacronym{MIMO}{MIMO}{multiple-input multiple-output}
\newacronym{MU}{MU}{multi-user}
\newacronym{OFDM}{OFDM}{orthogonal frequency-domain multiplexing}
\newacronym{IM}{IM}{index modulation}
\newacronym{IoT}{IoT}{Internet-of-Things}
\newacronym{QSM}{QSM}{quadrature spatial modulation}
\newacronym{BP}{BP}{belief propagation}
\newacronym{GaBP}{GaBP}{Gaussian belief propagation}
\newacronym{SM}{SM}{spatial modulation}
\newacronym{IQ}{IQ}{in-phase and quadrature}
\newacronym{ML}{ML}{maximum likelihood}
\newacronym{AP}{AP}{access point}
\newacronym{BER}{BER}{bit error rate}
\newacronym{P2P}{P2P}{point-to-point}
\newacronym{RCC}{RCC}{radar-communication coexistence}
\newacronym{CF-MIMO}{CF-MIMO}{cell-free MIMO}
\newacronym{AWGN}{AWGN}{additive white Gaussian noise}
\newacronym{SWIPT}{SWIPT}{simultaneous wireless information and power transfer}
\newacronym{GNSS}{GNSS}{global navigation satellite system}
\newacronym{CBTS}{CBTS}{consensus-based time synchronization}
\newacronym{PTP}{PTP}{precision time protocol}
\newacronym{CSI}{CSI}{channel state information}
\newacronym{MQAM}{$M$-QAM}{$M$-ary quadrature amplitude modulation}
\newacronym{IC}{IC}{interference cancellation}
\newacronym{SGA}{SGA}{scalar Gaussian approximation}
\newacronym{CLT}{CLT}{central limit theorem}
\newacronym{PDF}{PDF}{probability density function}
\newacronym{GB-ISTA}{GB-ISTA}{greedy boxed iterative soft-thresholding algorithm}
\newacronym{MP}{MP}{message passing}
\newacronym{SD}{SD}{sphere decoder}
\newacronym{FD}{FD}{full-duplex}
\newacronym{STC}{STC}{space-time coding}
\newacronym{SotA}{SotA}{state-of-the-art}
\newacronym{IER}{IER}{index vector error rate}
\newacronym{B5G}{B5G}{beyond fifth generation}
\newacronym{6G}{6G}{sixth generation}
\newacronym{STC-SM}{STC-SM}{space-time coded SM}
\newacronym{STC-QSM}{STC-QSM}{space-time coded QSM}
\newacronym{SC-IM}{SC-IM}{single-carrier IM}
\newacronym{SSK}{SSK}{space shift keying}
\newacronym{mmWave}{mmWave}{millimeter-wave}
\newacronym{THz}{THz}{Terahertz}
\newacronym{RIS}{RIS}{reflective intelligence surface}
\newacronym{RF}{RF}{radio frequency}
\newacronym{STBC}{STBC}{space-time block code}
\newacronym{MMSE}{MMSE}{minimum mean-squared-error}
\newacronym{CS}{CS}{compressive sensing}
\newacronym{i.i.d.}{i.i.d.}{independent and identically distributed}
\newacronym{JCAS}{JCAS}{joint communication and sensing}
\newacronym{ISAC}{ISAC}{integrated sensing and communication}
\newacronym{JCS}{JCS}{joint communication and sensing}
\newacronym{JRC}{JRC}{joint radar-communication}
\newacronym{JCR}{JCR}{joint communication-radar}
\newacronym{SE}{SE}{spectral efficiency}
\newacronym{EE}{EE}{energy efficiency}
\newacronym{UE}{UE}{user equipment}
\newacronym{OTFS}{OTFS}{orthogonal time frequency space}

\begin{document}
\title{\!\!\!Sparse\!\;\,Codesigned\!\;\,Communication\!\;\,and\!\;\,Radar\!\;\,Systems\! 
{\normalsize \textit{White paper for SPM Special Issue on "Signal Processing for the Integrated Sensing and Communications Revolution".}}}

\author{\normalsize Hyeon Seok Rou$^{1}$, Giuseppe Thadeu Freitas de Abreu$^{1}$, Saravanan Nagesh$^{2}$, Andreas Bathelt$^{2}$, \\ 
David González G.$^{3}$, Osvaldo Gonsa$^{3}$, and Hans-Ludwig Bloecher$^{4}$ \\[2ex]

\small $^1$School of Computer Science and Engineering, Constructor University, 28759 Bremen, Germany \\
$^2$Fraunhofer Institute for High Frequency Physics and Radar Techniques, 53343 Wachtberg, Germany \\
$^{3}$Wireless Communications Technologies, Continental AG, 65936 Frankfurt/Main, Germany \\
$^{4}$Autonomous Mobility, Continental A.D.C., 88131 Lindau, Germany \\
}

\maketitle


\section{\textbf{Short bio for each author}}
\par 
\textbf{Hyeon Seok Rou} [hrou@constructor.university] \textit{(Graduate Student Member, IEEE)} is a Ph.D. Candidate at Constructor University, Bremen, Germany, funded as a Research Associate at Continental AG on a researach project on 6G vehicular-to-everything (V2X) integrated sensing and communications (ISAC).
His research interests lie in the fields of ISAC, hyper-dimensional sparse modulation schemes, B5G/6G V2X wireless communications technology, and Bayesian inference. \\[-2ex]

\textbf{Giuseppe Thadeu Freitas de Abreu} [gabreu@constructor.university] \textit{(Senior Member, IEEE)} is a Full Professor of Electrical Engineering at Constructor University, Bremen, Germany. His research interests include communications theory, estimation theory, statistical modeling, wireless localization, cognitive radio, wireless security, MIMO systems, ultrawideband and millimeter wave communications, full-duplex and cognitive radio, compressive sensing, energy harvesting networks, random networks, connected vehicles networks, and many other topics. He has served as an editor for various IEEE Transactions, and currently serves as an editor to the IEEE Signal Processing Letters and the IEEE Communications Letters. 

\textbf{Saravanan Nagesh} [saravanan.nagesh@fhr.fraunhofer.de] is a Ph.D. Candidate with Prof. Dr.-Ing. Joachim Ender at the Fraunhofer Institute for High-Frequency Physics and Radar Techniques FHR in Wachtberg, Germany. His PhD is funded by the European Union’s, H2020-MSCA-ITN project MENELAOSNT. His research interest includes radar signal processing, waveform optimisation and multidimensional
sparse reconstruction.  \\[-2ex]

\textbf{Andreas Bathelt} [andreas.bathelt@fhr.fraunhofer.de] is a researcher at the Fraunhofer Institute for High Frequency Physics and Radar Techniques FHR. His research interests include integrated sensing and communications (ISAC), consensus-based time synchronization, compressed sensing, B5G/6G MIMO, and system identification, signal theory and estimation in the broader scope of radar operations including signal generation for and coordination of JRC capabilities.   \\[-2ex]

\textbf{David González G.} [david.gonzalez.g@ieee.org] \textit{(Senior Member, IEEE)}
is a Senior Research Engineer at Continental AG, Germany, and has previously served at Panasonic Research and Development Center, Germany.
His research interests include aspects of cellular networks and wireless communications, including interference management, radio access modeling and optimization, resource allocation, and vehicular communications. 
Since 2017, he has represented his last two companies as delegate in the 3GPP for 5G standardization, mainly focused on physical layer aspects and vehicular communications.  \\[-2ex]

\textbf{Osvaldo Gonsa} [osvaldo.gonsa@continental-corporation.com] is the Head of the Wireless Communications Technologies group by Continental AG in Frankfurt, Germany. 
He has worked in research and standardization in radio access network, serving as an advisor to the German Federal Ministry of Economy and Energy for the ``PAiCE" projects, and currently as the GSMA Advisory Board for automotive and the 6GKom project of the German Federal Ministry of Education and Research.  \\[-2ex]

\textbf{Hans-Ludwig Bloecher} [hans-ludwig.bloecher@continental-corporation.com] is the Head of Innovation Management of Business Area Autonomous Mobility, Continental A.D.C., Germany.
His areas of expertise include radar systems engineering, quasi optical radar and antenna components over 100GHz, ground/naval radars and multispectral missile seekers, 24/77/79 GHz UWB automotive radar platforms for ADAS/AD, in addition to contributions on regulatory, tandardization, platform integration and interference mitigation aspects related to automotive radar and UWB (ITU, CEPT, ETSI, DIN, beyond others).

\section{\textbf{History, motivation, and significance of the topic}}
\label{secrel}
\par 
The phenomenon of electromagnetic wave propagation, so brilliantly described in Maxwell's work \cite{Maxwell1865}, has been exploited in very distinct manners in the development of wireless communications and radar technologies.
Indeed, wireless communications systems built largely on information-theoretical foundations laid by Shannon \cite{ShannonBellLabs1948}, which allow for a significant degree of abstraction of the propagation phenomenon when modeling the relationship between signals and the information they carry.
In turn, despite several contributions on the utilization of information-theoretical principles in the design and analysis of radar systems \cite{BellTIT1993}, radar technology rests fundamentally on detection- and estimation-theoretical principles \cite{PoorBook1988, RichardsBook2005}, which rely on precise models of the employed signals and their interaction with the propagation environment.

Recently, however, driven by an ever-growing demand and increasingly limited spectral resources, wireless communications systems are quickly moving towards high-frequency channels, from  the \ac{mmWave} \cite{Wang_CST18} to \ac{THz} \cite{Song_TTHz22} bands.
Unlike the case of sub-6GHz channels, where the majority of currently deployed wireless communications systems operate, and which due to their highly dispersive nature can be sufficiently well characterized statistically \cite{Proakis2000, TseViswanath2008}, high-frequency channels models are significantly more descriptive, containing many parameters such as \ac{AoD}, \ac{AoA}, delays and Doppler shifts \cite{5GChannelModels}, which must be carefully taken into account in the design, and accurately estimated in the operation of higher frequency communication systems.

While a challenge in itself, this paradigm shift on role and form of channel modeling and estimation in wireless communication, is poised to narrow the functional gap between radar systems and \ac{B5G} or \ac{6G} communication systems, giving rise to the notion of \ac{ISAC}, also known as \ac{JCAS}.
In fact, it is envisioned that \ac{B5G}/\ac{6G} wireless systems will strongly rely on \ac{mMIMO} technology to mitigate challenges of the \ac{mmWave}/\ac{THz} channel, further empowering those systems to accurately perform environment sensing and target tracking functions previously delivered by radar systems, thus enabling new applications in areas such as autonomous vehicles, drone networks, and smart cities \cite{PinTan_JCS21, Rappaport_Access19}.

It is important to recognize, however, that no matter how accurate the radar-like sensing functions that future communication systems become able to provide, other fundamental requirements of radar applications, such as low latency, high reliability and flexible power budget, ensure that a demand for radar systems that operate independent of eventual co-existing communication systems will always exist \cite{PatoleSPMag2017, PetrovWCMag2019}.
We emphasize that the latter is not a contradiction to the concept of \ac{ISAC}, but merely a recognition of the fact that the integration of communications and radar functions \cite{Liu_2023} can (and must) be accomplished in distinct manners, ranging from coexistence \cite{Zheng_SPM19} to full incorporation \cite{NiISWCS2022}.
Coincidentally, literature also recognizes the aforementioned fact by classifying \cite{Zhang_JSTSP21,Liu_TC20} existing contributions fundamentally as either communication-centric or radar-centric schemes.

In particular, schemes that aim to achieve sensing functionality via the estimation of environment parameters of the propagation model corresponding to signals designed for a given communication system, are classified as communication-centric.
Examples of such an approach can be found based on various waveforms such as \ac{OFDM} \cite{Liu_RADAR16}, IEEE 802.11ad \cite{Kumari_TVT18}, \ac{OTFS} \cite{Raviteja_Radarconf19} and Wi-Fi \cite{Tan_CM18}.

On the other hand, radar-centric schemes are deemed to be those which aim to add communication functionality to a radar systems, either by modulating information on top of the sensing chirp waveforms over the time-, frequency-, or code-domain \cite{Gaglione_Radarconf16,Saddik_TMTT07}, or by leveraging \ac{IM}-based techniques \cite{Ma_JSTSP21,Huang_TSP20}.

While the communications- versus radar-centric classification is proper in describing the status quo, it is not as useful in providing insight on how the \ac{ISAC} area is likely to develop in the future.
Indeed, there are already recent contributions which cannot be strictly classified as either communication- or radar-centric, since their novelty lies in either the design of new waveforms, optimized to achieve both functionalities \cite{Yuan_TVT21, Liu_CL17}, or in the approach to model the propagation phenomenon in a manner agnostic to the waveform employed \cite{Rou_Asilomar22_JCAS,Rou_arxiv23_ISAC}.

In view of all the above, this short article is a brief description of an intended contribution offering an alternative perspective of \ac{ISAC} systems, whereby the integration of radar and communications functionalities is achieved via the design of either the signals or the resource utilization mechanisms (or both), such as to enable both radar and communication functionalities.
Within that perspective, we furthermore focus on methods that exploit sparsity in the aforementioned design, which is again to be understood both in terms of resource utilization and the waveforms themselves.

\vspace{1ex}
\textbf{Intended contributions of the full article include}: \vspace{-0.25ex}
\begin{itemize}
\item A thorough introduction of \ac{ISAC} systems with a focus on techniques leveraging sparsity in resource and signal processing;
\item A \ac{SotA} survey of existent \ac{ISAC} techniques, categorizing the various methods via their signal design and resource utilization properties and comparing the design, behavior, and challenges of each method;
\item Examples of novel \ac{SCCR} techniques, which leverage the \ac{SM} framework and sparse radar techniques to co-design the antenna allocation pattern codebooks for each subsystem;
\item A motivation and potential direction of research for multi-static synchronization and network sensing scenarios.
\end{itemize}

\section{\textbf{Outline of the proposed SI paper}}
\label{secpr}

In addition to the introduction and the significance of the proposed topic as motivated in the previous section, we aim to further deliver the contributions outlined in the following.

\subsection{\textbf{State-of-the-Art Towards Sparse Codesigned Communication and Radar (SCCR)}}

In this section, we aim to provide a discussion on the various \ac{SotA} \ac{ISAC} techniques motivating the proposed \ac{SCCR} design.

\vspace{1ex}
\subsubsection{Static Resource Allocation Codesign} $~$ \vspace{0.25ex}

A very simple yet straight-forward method to realize the coexistence of radar and communication functions in the same system is to simply share the available resources, via a subdivision and static allocation of subcarriers, antennas, time slots and so on, to the respective subsystems.

The rigid implementation and design of such schemes limit, however, the degrees-of-freedom available for performance enhancement through system design, because optimal resource allocation (typically established with the aim of diminishing interference between the two subsystems) must be performed in average terms \cite{Zheng_JSTSP18, Liu_TSP18}, which inherently imply a decreased spectral efficiency and diversity for the communication subsystem, and decreased differentiability for the radar subsystem, due to the reduced aperture and undesirable sidelobe behavior.

\vspace{1ex}
\subsubsection{Dynamic Resource Allocation Codesign} $~$ \vspace{0.25ex}

In light of the aforementioned bottleneck, which cannot be compensated by either massification, due to exponentially growing costs, or by static methods, due to the reduced amount of resource per subsystem \cite{Chiriyath_TCCN17}, the sparsification-based \ac{ISAC} approach, here referred to as the \acf{SCCR}, emerges as a possible solution.
For instance, the introduction of sparsity in the resource domain allows for the exploitation of \ac{IM}, which are known to achieve higher energy and spectral efficiency compared to the dense counterpart \cite{Sugiura_Access17}.

This trend can be observed in recently-proposed systems in which communications and radar functions are \textit{interleaved}, an example of which is \cite{Ma_TVT21}, where signals are sparsely distributed in a \acs{MIMO} array by using \ac{GSM} \cite{Younis_Asilomar10}.
Unfortunately, the interleaving approach exhibits significant performance gaps compared to the corresponding full-array techniques, which indicates that further steps to codesign the radar and communication waveforms is necessary, as discussed in the sequel.

\subsection{\textbf{Sparse Codesigned Communication and Radar (SCCR) Design}}

Given all the above, we aim in the full manuscript to discuss the potential approaches whereby not only the resource domain of the transmit signal, but also the signals themselves are designed in a sparse manner, which corresponds to a type of \ac{SCCR} scheme with ``dynamic signal and resource utilization codesign''.
To elaborate a little further, while the notion of sparse signaling is common in communications systems, it is less common (but not unheard of) in the domain of radar signal design.
To understand the reason, consider that an ideal chirp (the theoretically perfect radar signal) is indeed a continuous waveform spanning a certain frequency range, and therefore associated with a flat spectrum, while a sparsified chirp ($i.e.$, a chirp with gaps) has a more complex spectrum.

In conjunction with a carefully designed sparsified utilization of resources, however, the desired spectral features of radar signals can be restored \cite{Larsson_TSP02, Karlsson_TAES14}, {\bf which in turn makes the case for a \underline{codesigned} sparsification of resources and signals jointly, as here proposed}.
Under the aforementioned argument, the full article will elaborate on such an approach.
Furthermore, two novel \ac{SCCR} techniques with the dynamic signal and resource utilization codesign will be described as examples to illustrate the effectiveness of the \ac{SCCR} scheme, which are very briefly described below, in addition to a final remark on the extension of the proposed design in a multi-static, cooperative network sensing scenario incorporating the key synchronization challenge.

\vspace{1ex}
\subsubsection{Adaptive Discretized Array Approach} $~$ \vspace{0.25ex}

In the first proposed design method, the full available radar array aperture is discretized (quantized) into an effective array with the same width (aperture) but a smaller number of intermediate antenna elements.
The reduced discretization therefore yields sporadic antennas in the array, that consist of unused elements by the radar system, which can be corresponidngly allocated for the transmission of communication signals. 

Consequently, the discretized nature of the radar aperture in \ac{MIMO} radar systems gives rise to their inherent angular resolution, which is contingent upon the quantity of antenna elements employed. 
Augmenting the number of elements in the array directly engenders an amelioration in angular resolution, as demonstrated in \cite{4350230}.
However, it was recently shown that achieving comparable angular resolution while employing a reduced number of antenna elements is plausible, provided that a fixed boresight is upheld. 
Pertinent investigations in this domain are elaborated upon in \cite{9103001} and \cite{8904530}, focusing on sparse array architectures. 
These studies furnish a conceptual underpinning for realizing efficient angular resolution within our specified system context.

Furthermore, due to the potential non-uniform arrangement of elements across the aperture arising during the discretization, unintended sidelobes manifest as artifacts, potentially obscuring detection targets. 
A promising avenue to mitigate these sidelobes involves treating the quandary as an under-determined problem that can be resolved via the implementation of sparse reconstruction algorithms, as expounded in \cite{9904978}.

Therefore, this proposed methodology is anticipated to showcase heightened angular resolution even in the presence of a diminished complement of radar resources, a consequence of the maintained aperture size, while upholding desirable sidelobe behaviors.
Moreover, by modulating the degree and the pattern of the discretization, the proposed scheme will aim to dynamically allocate performance gains to either align with specific communication quality-of-service requisites or to fulfill desired radar sensing resolutions.

\vspace{1ex}
\subsubsection{Sparse Radar Waveform Recovery Approach} $~$ \vspace{0.25ex}

In the second proposed scheme, the totally available resources are first fully densely occupied with optimized radar signals, as to reap the maximum radar sensing performance in the utilized array.
Then, an independently modulated sparsely-distributed \ac{IM}-based communication symbols are overlaid and \textit{replaces} the already occupied resources, replacing the radar signal wherever, effectively nullifying the disrupting radar signal spectrum.
The objective of this approach would consequently be to enable robustness in the processing the received radar signals with effective ``blockages'', such that there is minimal performance degradation even with the unavailable resources in the signal and spectrum domain, alike to a data completion approach.

A predecessor of such design can be found in \cite{Guha.2022} where a compressed sensing problem is formulated to bridge gaps in the spectrum of a radar signal, which had not been explicitly considered in previous works \cite{Larsson_TSP02, Karlsson_TAES14}. 
By borrowing the concept of fusion frames \cite{Aceska.2018}, and considering the fixed sensing matrix structure of radar signals, the effective frequency gaps become missing block rows of the  matrix.
By sub-dividing the sensing matrix into sub-matrices, the distances between columns of the sub-matrices increase to retain the solvability of the estimation problem even in cases of large unavailable gaps. 
The high resolution range estimation is acquired by fusing the individual results based on the jointly-defined range support. 
Extending the above single-antenna results for MIMO systems, a significant opportunity is that the gap in a signal on one antenna does not necessarily coincide with the gap in the signals of other antennas. 
The estimation problem is then required to recover the gaps based on the signals of other antennas and jointly consider all individual signals over all antennas, which was indeed recently hinted in \cite{Bathelt.2023} through a distributed-band matrix approach.

In summary, the scheme will therefore effectively result in a pure \ac{IM} communication subsystem domain, and a sparsely punctured radar subsystem domain, which via sparse radar signal processing, can be accurately reconstructed to minimize the effect of the lost radar signals in the estimation performance.

\vspace{1ex}
\subsubsection{Synchronization for Networked Sensing} $~$\vspace{0.25ex}

Finally, we will also aim to provide a discussion on the extension of the mono-static scenario discussed above and for most of the \ac{SotA}, into a multi-static environment with an increased number of receiver nodes in the network, which will cooperatively localize and track the targets.

In the context of networked \ac{ISAC}, synchronization is a critical building block.
However, common approaches may become unsuitable due to the concomitant costs of atomic-clocks or \ac{GNSS}-disciplined clocks, or the problems in case of an outage of the master of master-slave protocols, as in \ac{PTP} \cite{IEEEInstrumentationandMeasurementSociety.2008}.
Considering that sensing fundamentally only needs an agreement on some time basis, the distributed, multi-agent approach of \ac{CBTS} \cite{Tian.2016} becomes a viable approach, even more so as the required message exchange is already inherently given by communication network. 
These algorithms are based on static consensus algorithms \cite{OlfatiSaber.2007} and bring the parameters, assumed to be time-invariant, of the local clocks into agreement. 
Another approach currently developed by the authors is the use of dynamic consensus algorithms \cite{Kia.2019}, where the consensus is achieved by bringing the clock values directly into agreement. 
This allows for the compensation of the inevitable variations of the clock oscillators. 
In light of the above the full article will give an outline of the existing approaches and discuss the ongoing development based on such dynamic consensus methods to enable efficient, high performance networked \ac{ISAC}.



\begin{thebibliography}{10}
\providecommand{\url}[1]{#1}
\csname url@samestyle\endcsname
\providecommand{\newblock}{\relax}
\providecommand{\bibinfo}[2]{#2}
\providecommand{\BIBentrySTDinterwordspacing}{\spaceskip=0pt\relax}
\providecommand{\BIBentryALTinterwordstretchfactor}{4}
\providecommand{\BIBentryALTinterwordspacing}{\spaceskip=\fontdimen2\font plus
\BIBentryALTinterwordstretchfactor\fontdimen3\font minus
\fontdimen4\font\relax}
\providecommand{\BIBforeignlanguage}[2]{{%
\expandafter\ifx\csname l@#1\endcsname\relax
\typeout{** WARNING: IEEEtran.bst: No hyphenation pattern has been}%
\typeout{** loaded for the language `#1'. Using the pattern for}%
\typeout{** the default language instead.}%
\else
\language=\csname l@#1\endcsname
\fi
#2}}
\providecommand{\BIBdecl}{\relax}
\BIBdecl

\bibitem{Maxwell1865}
J.~C. Maxwell, ``A dynamical theory of the electromagnetic field,''
\emph{Philosophical Trans. of the Royal Society of London}, vol. 155,
pp. 459--512, 1865.

\bibitem{ShannonBellLabs1948}
C.~E. Shannon, ``A mathematical theory of communication,'' \emph{Bell Syst.
Tech. J.}, vol.~27, pp. 379 -- 423, 623 -- 656, Jul.-Oct. 1948.

\bibitem{BellTIT1993}
M.~Bell, ``Information theory and radar waveform design,'' \emph{IEEE
Transactions on Information Theory}, vol.~39, no.~5, pp. 1578--1597, 1993.

\bibitem{PoorBook1988}
H.~V. Poor, \emph{An Introduction to Signal Detection and Estimation}.\hskip
1em plus 0.5em minus 0.4em\relax Springer, 1988.

\bibitem{RichardsBook2005}
\BIBentryALTinterwordspacing
M.~A. Richards, \emph{Fundamentals of Radar Signal Processing}.\hskip 1em plus
0.5em minus 0.4em\relax <country>US</country>: McGraw-Hill Professional,
2005.
\BIBentrySTDinterwordspacing

\bibitem{Wang_CST18}
X.~Wang, L.~Kong, F.~Kong, F.~Qiu, M.~Xia, S.~Arnon, and G.~Chen, ``Millimeter
wave communication: A comprehensive survey,'' \emph{IEEE Commun. Surveys
Tuts.}, vol.~20, no.~3, pp. 1616--1653, 2018.

\bibitem{Song_TTHz22}
H.-J. Song and N.~Lee, ``Terahertz communications: Challenges in the next
decade,'' \emph{IEEE Trans. THz Sci. Technol.}, vol.~12, no.~2, pp. 105--117,
2022.

\bibitem{Proakis2000}
J.~G. Proakis, \emph{Digital Communications, Fourth Edition}.\hskip 1em plus
0.5em minus 0.4em\relax New York, NY: Mc-Graw-Hill, 2000.

\bibitem{TseViswanath2008}
D.~Tse and P.~Viswanath, \emph{Fundamentals of Wireless Communications}.\hskip
1em plus 0.5em minus 0.4em\relax Cambridge University Press, 2008.

\bibitem{5GChannelModels}
\BIBentryALTinterwordspacing
``{5G} channel model for bands up to 100 {GHz},'' 2015.
\BIBentrySTDinterwordspacing

\bibitem{PinTan_JCS21}
D.~K. Pin~Tan \emph{et al.,}
``Integrated sensing and communication in 6g: Motivations, use cases,
requirements, challenges and future directions,'' in \emph{2021 1st IEEE
International Online Symposium on Joint Communications \& Sensing (JC\&S)}, 2021, pp. 1--6.

\bibitem{Rappaport_Access19}
T.~S. Rappaport, Y.~Xing, O.~Kanhere, S.~Ju, A.~Madanayake, S.~Mandal,
A.~Alkhateeb, and G.~C. Trichopoulos, ``Wireless communications and
applications above 100 {GHz}: Opportunities and challenges for {6G} and
beyond,'' \emph{IEEE Access}, vol.~7, pp. 78\,729--78\,757, 2019.

\bibitem{PatoleSPMag2017}
S.~M. Patole, M.~Torlak, D.~Wang, and M.~Ali, ``Automotive radars: A review of
signal processing techniques,'' \emph{IEEE Signal Processing Magazine},
vol.~34, no.~2, pp. 22--35, 2017.

\bibitem{PetrovWCMag2019}
V.~Petrov, G.~Fodor, J.~Kokkoniemi, D.~Moltchanov, J.~Lehtomaki, S.~Andreev,
Y.~Koucheryavy, M.~Juntti, and M.~Valkama, ``On unified vehicular
communications and radar sensing in millimeter-wave and low terahertz
bands,'' \emph{IEEE Wireless Communications}, vol.~26, no.~3, pp. 146--153,
2019.

\bibitem{Liu_2023}
\BIBentryALTinterwordspacing
F.~Liu, L.~Zheng, Y.~Cui, C.~Masouros, A.~P. Petropulu, H.~Griffiths, and Y.~C.
Eldar, ``Seventy years of radar and communications: The road from separation
to integration,'' \emph{{IEEE} Signal Processing Magazine}, vol.~40, no.~5,
pp. 106--121, jul 2023. [Online]. Available:
\url{https://doi.org/10.1109%2Fmsp.2023.3272881}
\BIBentrySTDinterwordspacing

\bibitem{Zheng_SPM19}
L.~Zheng, M.~Lops, Y.~C. Eldar, and X.~Wang, ``Radar and communication
coexistence: An overview: A review of recent methods,'' \emph{IEEE Signal
Process. Mag.}, vol.~36, no.~5, pp. 85--99, 2019.

\bibitem{NiISWCS2022}
Y.~Ni, Z.~Wang, P.~Yuan, and Q.~Huang, ``An {AFDM}-based integrated sensing and
communications,'' in \emph{2022 International Symposium on Wireless
Communication Systems (ISWCS)}, 2022, pp. 1--6.

\bibitem{Zhang_JSTSP21}
J.~A. Zhang, F.~Liu, C.~Masouros, R.~W. Heath, Z.~Feng, L.~Zheng, and
A.~Petropulu, ``An overview of signal processing techniques for joint
communication and radar sensing,'' \emph{IEEE J. Sel. Topics Sig. Proc.},
vol.~15, no.~6, pp. 1295--1315, 2021.

\bibitem{Liu_TC20}
F.~\emph{et al.}, ``Joint radar and communication design: Applications,
state-of-the-art, and the road ahead,'' \emph{IEEE Trans. Commun.}, vol.~68, no.~2, 2020.

\bibitem{Liu_RADAR16}
Y.~Liu, G.~Liao, and Z.~Yang, ``Range and angle estimation for {MIMO-OFDM}
integrated radar and communication systems,'' in \emph{2016 CIE International
Conference on Radar (RADAR)}, 2016, pp. 1--4.

\bibitem{Kumari_TVT18}
{P. Kumari, J. Choi, N. Gonz{\'a}lez-Prelcic and R. W. Heath}, ``{IEEE}
802.11ad-based radar: An approach to joint vehicular communication-radar
system,'' \emph{IEEE Trans. Veh. Technol.}, vol.~67, no.~4, pp. 3012--3027,
2018.

\bibitem{Raviteja_Radarconf19}
P.~Raviteja, K.~T. Phan, Y.~Hong, and E.~Viterbo, ``Orthogonal time frequency
space ({OTFS}) modulation based radar system,'' in \emph{2019 IEEE Radar
Conference (RadarConf)}, 2019, pp. 1--6.

\bibitem{Tan_CM18}
B.~Tan, Q.~Chen, K.~Chetty, K.~Woodbridge, W.~Li, and R.~Piechocki,
``Exploiting {WiFi} channel state information for residential healthcare
informatics,'' \emph{IEEE Commun. Mag.}, vol.~56, no.~5, pp. 130--137, 2018.

\bibitem{Gaglione_Radarconf16}
D.~Gaglione, C.~Clemente, C.~V. Ilioudis, A.~R. Persico, I.~K. Proudler, and
J.~J. Soraghan, ``Fractional fourier based waveform for a joint
radar-communication system,'' in \emph{2016 IEEE Radar Conference
(RadarConf)}, 2016, pp. 1--6.

\bibitem{Saddik_TMTT07}
G.~N. Saddik, R.~S. Singh, and E.~R. Brown, ``Ultra-wideband multifunctional
communications/radar system,'' \emph{IEEE Transactions on Microwave Theory
and Techniques}, vol.~55, no.~7, pp. 1431--1437, 2007.

\bibitem{Ma_JSTSP21}
D.~Ma, N.~Shlezinger, T.~Huang, Y.~Liu, and Y.~C. Eldar, ``{FRaC}: {FMCW}-based
joint radar-communications system via index modulation,'' \emph{IEEE J. Sel.
Topics Signal Process.}, vol.~15, no.~6, pp. 1348--1364, 2021.

\bibitem{Huang_TSP20}
T.~Huang, N.~Shlezinger, X.~Xu, Y.~Liu, and Y.~C. Eldar, ``{MAJoRCom}: A
dual-function radar communication system using index modulation,'' \emph{IEEE
Trans. on Signal Process.}, vol.~68, pp. 3423--3438, 2020.

\bibitem{Yuan_TVT21}
{X. Yuan \emph{et al.}}, ``Spatio-temporal power optimization for {MIMO} joint
communication and radio sensing systems with training overhead,'' \emph{IEEE
Trans. Veh. Technol.}, vol.~70, no.~1, pp. 514--528, 2021.

\bibitem{Liu_CL17}
Liu .~\emph{et al.}, ``Adaptive \ac{OFDM} integrated radar and communications waveform
design based on information theory,'' \emph{IEEE Commun. Let.}, vol.~21,
no.~10, pp. 2174--2177, 2017.

\bibitem{Rou_Asilomar22_JCAS}
H.~S. Rou, G.~T. Freitas~de Abreu, D.~G. G, and O.~Gonsa, ``Asymmetric bilinear
inference for joint communications and environment sensing,'' in \emph{2022
56th Asilomar Conference on Signals, Systems, and Computers}, 2022, pp.
1111--1115.

\bibitem{Rou_arxiv23_ISAC}
------, ``Integrated sensing and communications for {3D} object imaging via
bilinear inference,'' \emph{arXiv preprint arXiv:2308.10423}, Jul. 2023.

\bibitem{Zheng_JSTSP18}
L.~Zheng, M.~Lops, and X.~Wang, ``Adaptive interference removal for
uncoordinated radar/communication coexistence,'' \emph{IEEE Journal of
Selected Topics in Signal Processing}, vol.~12, no.~1, pp. 45--60, 2018.

\bibitem{Liu_TSP18}
F.~Liu, C.~Masouros, A.~Li, T.~Ratnarajah, and J.~Zhou, ``{MIMO} radar and
cellular coexistence: A power-efficient approach enabled by interference
exploitation,'' \emph{IEEE Trans. Signal Process.}, vol.~66, no.~14, pp.
3681--3695, 2018.

\bibitem{Chiriyath_TCCN17}
A.~R. Chiriyath, B.~Paul, and D.~W. Bliss, ``Radar-communications convergence:
Coexistence, cooperation, and co-design,'' \emph{IEEE Transactions on
Cognitive Communications and Networking}, vol.~3, no.~1, pp. 1--12, 2017.

\bibitem{Sugiura_Access17}
S.~Sugiura, T.~Ishihara, and M.~Nakao, ``State-of-the-art design of index
modulation in the space, time, and frequency domains: Benefits and
fundamental limitations,'' \emph{IEEE Access}, vol.~5, pp. 21\,774--21\,790,
2017.

\bibitem{Ma_TVT21}
D.~Ma, N.~Shlezinger, T.~Huang, Y.~Shavit, M.~Namer, Y.~Liu, and Y.~C. Eldar,
``Spatial modulation for joint radar-communications systems: Design,
analysis, and hardware prototype,'' \emph{IEEE Transactions on Vehicular
Technology}, vol.~70, no.~3, pp. 2283--2298, 2021.

\bibitem{Younis_Asilomar10}
{A. Younis} \emph{et al.,}, ``Generalised spatial
modulation,'' in \emph{Proc. 44th Asilomar Conf. Signals, Syst. Comput.},
2010, pp. 1498--1502.

\bibitem{Larsson_TSP02}
E.~Larsson, P.~Stoica, and J.~Li, ``Amplitude spectrum estimation for
two-dimensional gapped data,'' \emph{IEEE Trans. on Sig. Proc.},
vol.~50, no.~6, pp. 1343--1354, June 2002.

\bibitem{Karlsson_TAES14}
J.~{Karlsson}, W.~{Rowe}, L.~{Xu}, G.~{Glentis}, and J.~{Li}, ``Fast
missing-data {IAA} with application to notched spectrum {SAR},'' \emph{IEEE
Transactions on Aerospace and Electronic Systems}, vol.~50, no.~13, pp.
959--971, 2014.

\bibitem{4350230}
J.~Li and P.~Stoica, ``{MIMO} radar with colocated antennas,'' \emph{IEEE Signal
Processing Magazine}, vol.~24, no.~5, pp. 106--114, 2007.

\bibitem{9103001}
A.~Ajorloo, A.~Amini, E.~Tohidi, M.~H. Bastani, and G.~Leus, ``Antenna
placement in a compressive sensing-based colocated {MIMO} radar,'' \emph{IEEE
Transactions on Aerospace and Electronic Systems}, vol.~56, no.~6, pp.
4606--4614, 2020.

\bibitem{8904530}
D.~Mateos-Núñez, M.~A. González-Huici, R.~Simoni, F.~B. Khalid,
M.~Eschbaumer, and A.~Roger, ``Sparse array design for automotive mimo
radar,'' in \emph{2019 16th European Radar Conference (EuRAD)}, 2019, pp.
249--252.

\bibitem{9904978}
S.~Nagesh, J.~Ender, and M.~A. González-Huici, ``Array position optimisation
for compressed sensing mimo radar based on mutual coherence minimisation,''
in \emph{2022 23rd International Radar Symposium (IRS)}, 2022, pp. 98--103.

\bibitem{Guha.2022}
S.~Guha \emph{et al.}, ``Radar band fusion using
frame-based compressed sensing,'' \emph{IEEE Journal of Selected Topics in
Signal Processing}, pp. 1--13, 2022.

\bibitem{Aceska.2018}
R.~Aceska, J.-L. Bouchot, and S.~Li, ``Fusion frames and distributed
sparsity,'' in \emph{Frames and Harmonic Analysis}, ser. Contemporary
Mathematics Ser, Y.~Kim, S.~K. Narayan, and G.~Picioroaga, Eds.\hskip 1em
plus 0.5em minus 0.4em\relax Providence: {American Mathematical Society},
2018, vol. 706, pp. 47--61.

\bibitem{Bathelt.2023}
A.~Bathelt and R.~Thill, ``Radar-sensing based on non-contiguous {OFDM} signalsusing compressed sensing,'' in \emph{Proc. of the EuRAD}, 2023.

\bibitem{IEEEInstrumentationandMeasurementSociety.2008}
\BIBentryALTinterwordspacing
{IEEE Instrumentation and Measurement Society} and {IEEE-SA Standards Board},
``{IEEE} standard for a precision clock synchronization protocol for
networked measurement and control systems,'' New York, N.Y., July 24, 2008.
\BIBentrySTDinterwordspacing

\bibitem{Tian.2016}
Y.-P. Tian, S.~Zong, and Q.~Cao, ``Structural modeling and convergence analysis
of consensus-based time synchronization algorithms over networks:
Non-topological conditions,'' \emph{Automatica}, vol.~65, pp. 64--75, 2016.

\bibitem{OlfatiSaber.2007}
R.~Olfati-Saber \emph{et al.}, ``Consensus and cooperation in
networked multi-agent systems,'' \emph{Proceedings of the IEEE}, vol.~95,
no.~1, pp. 215--233, 2007.

\bibitem{Kia.2019}
S.~S. Kia, B.~{van Scoy}, J.~Cortes, R.~A. Freeman, K.~M. Lynch, and
S.~Martinez, ``Tutorial on dynamic average consensus: The problem, its
applications, and the algorithms,'' \emph{IEEE Control Systems}, vol.~39,
no.~3, pp. 40--72, 2019.

\end{thebibliography}

\end{document}